\documentclass{aastex63}

\usepackage[T1]{fontenc}
\usepackage{bm}        % for math
\usepackage{amssymb, amsmath}   % for math
\usepackage{cancel}
\usepackage{color}

\DeclareMathAlphabet{\mathsfit}{\encodingdefault}{\sfdefault}{m}{sl}
\SetMathAlphabet{\mathsfit}{bold}{\encodingdefault}{\sfdefault}{bx}{sl}
\shorttitle{PSP SEP results}
\shortauthors{Guo et al.}
\begin{document}

\title{Variable Ion Compositions of Solar Energetic Particle Events in the Inner Heliosphere: \\A Field-line Braiding Model with Compound Injections}%\\A Field-line Braiding Model with Compound Injections for Variable Ion Compositions of Solar Energetic Particles Events in the Inner Heliosphere}

\correspondingauthor{Fan Guo}
\email{guofan@lanl.gov}

%\author[0000-0001-9826-1759]{Haocheng Zhang}
%\affiliation{Department of Physics and Astronomy \\
%Purdue University \\
%West Lafayette, IN 47907, USA}
%\affiliation{New Mexico Consortium \\
%Los Alamos, NM 87544, USA}

%\author[0000-0003-1503-2446]{Dimitrios Giannios}
%\affiliation{Department of Physics and Astronomy \\
%Purdue University \\
%West Lafayette, IN 47907, USA}

\author[0000-0003-4315-3755]{Fan Guo}
\affiliation{Theoretical Division,
Los Alamos National Laboratory,
Los Alamos, NM 87545, USA}

\author[0000-0003-3936-5288]{Lulu Zhao}
\affiliation{Department of Climate and Space Sciences and Engineering, University of Michigan, Ann Arbor, MI, 48109, USA}

\author[0000-0002-0978-8127]{Christina M. S. Cohen}
\affiliation{California Institute of Technology, Pasadena, CA 91125, USA}

\author[0000-0002-0850-4233]{Joe Giacalone}
\affiliation{University of Arizona, Tucson, AZ 85721, USA}

\author[0000-0002-0156-2414]{R. A. Leske}
\affiliation{California Institute of Technology, Pasadena, CA 91125, USA}

\author[0000-0002-2825-3128]{M. E. Wiedenbeck}
\affiliation{Jet Propulsion Laboratory, California Institute of Technology, Pasadena, CA 91109, USA}

\author[0000-0002-2825-3128]{S. W. Kahler}
\affiliation{Air Force Research Laboratory, Space Vehicles Directorate, 3550 Aberdeen Ave., Kirtland AFB, NM 87117, USA}

\author[0000-0001-5278-8029]{Xiaocan Li}
\affiliation{Dartmouth College, Hanover, NH 03750, USA}

\author[0000-0002-9504-641X]{Qile Zhang}
\affiliation{Theoretical Division,
Los Alamos National Laboratory,
Los Alamos, NM 87545, USA}

\author[0000-0003-1093-2066]{George C. Ho}
\affiliation{The Johns Hopkins University, Applied Physics Laboratory, Laurel MD 20723, USA}

\author[0000-0002-7318-6008]{Mihir I. Desai}
\affiliation{Department of Physics and Astronomy, University of Texas at San Antonio, San Antonio, TX 78249, USA}
\affiliation{Southwest Research Institute, San Antonio, TX 78249, USA}

%\affiliation{New Mexico Consortium \\
%Los Alamos, NM 87544, USA}
%\author[]{Lulu Zhao}
%\affiliation{xx}

\begin{abstract}
We propose a model for interpreting highly variable ion composition ratios in solar energetic particles (SEP) events recently observed by Parker Solar Probe (PSP) at $0.3 - 0.45$ astronomical unit. We use numerical simulations to calculate SEP propagation in a turbulent interplanetary magnetic field with a Kolmogorov power spectrum from large scale down to the gyration scale of energetic particles. We show that when the source regions of different species are offset by a distance comparable to the size of the source regions, the observed energetic particle composition He/H can be strongly variable over more than two orders of magnitude, even if the source ratio is at the nominal value. Assuming a $^3$He/$^4$He source ratio of $10 \%$ in impulsive $^3$He-rich events and the same spatial offset of the source regions, the $^3$He/$^4$He ratio at observation sites also vary considerably. The variability of the ion composition ratios depends on the radial distance, which can be tested by observations made at different radial locations. We discuss the implication of these results on the variability of ion composition of impulsive events and on further PSP and Solar Orbiter observations close to the Sun.
\end{abstract}

\keywords{Solar energetic particles, Interplanetary turbulence, Interplanetary magnetic fields}


\section{Introduction} \label{sec:intro}

The acceleration and transport of energetic charged particles is a remarkable problem in heliophysics and astrophysics. While much progress has been made, many aspects of observations remain mysterious. Different acceleration and transport effects are often entangled when energetic particles are observed, making it challenging to distinguish these effects and evaluate their relative importance. The Parker Solar Probe (PSP) and Solar Orbiter (SO) provide unique opportunities for studying energetic particles closer to the Sun, where interplanetary scattering, among other effects, is substantially reduced compared to observations made at 1 AU \citep{Desai2016,McComas2016,Rodrguez2020}. 

During the first several orbits, PSP has observed a number of small Solar Energetic Particle (SEP) events at $0.17$ - $0.5$ AU. These have provided much insight yet present new puzzles to particle acceleration and transport processes. Especially, \citet{Cohen2021} reported that the helium to hydrogen ratio (He/H)\footnote{We refer $^4$He as He throughout the paper unless we specifically discuss their isotopes.} showed substantial event-to-event variation, $0.33\%\pm0.13\%$, $1.6\%\pm0.9$\% and $17.6\%\pm4.7\%$ for three SEP events originating from the same active region, whereas the nominal ratio measured at 1 AU is about $3.6\%$ \citep{Reames1995}. An earlier SEP event showed a He/H ratio of $\sim5.2\%$ \citep{Leske2020}. Variations of $^3$He/$^4$He, O/He, electron/proton and other heavy ions are also evident in the first SEP events observed by PSP \citep{Leske2020,Wiedenbeck2020,Cohen2021}.  Using measurements at 1 AU, \citet{Reames2019} reported a reduced He/C ratio and possibly He/H ratio by more than one order of magnitude in some events. It was also noted that these helium-poor events are among the smaller impulsive events, indicating an event-size dependent variation. \citet{Kahler2021} reported gradual SEP events that show large variations of He/H ratios. They find that the He/H ratios in the gradual SEP events are correlated with that in the solar wind.  While \citet{Reames2019} suggested that these He-poor events are due to the first ionization potential (FIP), it is interesting to note that some of the reported events have He/C up to four times of the ``nominal'' ratio in impulsive SEP events \citep{Reames1995}. The FIP effect may generate a He reduction by several times \citep{Laming2021}, but has difficulties in explaining the variation of He/H over nearly two orders of magnitude observed by PSP. A relevant observation by Advanced Composition Explorer (ACE) showed that in some events the $^{22}$Ne/$^{20}$Ne ratio -- with nearly the same charge to mass ratio -- has a three-fold event-to-event variation (0.7-2 times of the solar wind value) \citep{Leske1999b}. \citet{Wiedenbeck2008} have reported the variation over a number of heavy ion species and noted that heavy ion isotope ratios (e.g., $^{26}$Mg/$^{24}$Mg and $^{22}$Ne/$^{20}$Ne) are more variable in $^3$He-rich events than in large gradual SEP events. Together, these observations suggested a systematic variability over a wide range of species, which has not been explored much in either observation or theory \citep{Leske1999a,Mason1979}. The ongoing PSP and SO missions provide a unique opportunity to study the variable ion composition in SEP events. Here we propose a scenario, where the observed variability of ion composition is due to the propagation of SEPs in turbulent interplanetary magnetic fields from compound injections at the source regions.

It has been well established that the transport of energetic particles in the inner heliosphere is governed by magnetic turbulence \citep{Jokipii1971,Zank2014}. %Understanding the influence of magnetic turbulence on the motions of charged particles is essential to model the transport of energetic charged particles in the heliosphere. 
At small scales, particles resonantly interact with turbulence and get scattered in their pitch angles \citep{Jokipii1966,Jokipii1971}, whereas at large scales, magnetic field lines of force braid and meander, leading to a transverse diffusion to the mean magnetic field \citep{Jokipii1969,Giacalone1999,Matthaeus2003}. 
%Examining the processes in different scales is important for understanding the observations of SEPs. 
Observations of the dropouts, namely sharp changes of energetic particle fluxes simultaneous at all energies, indicate diffusion transverse to the local magnetic field is weak \citep{Mazur2000,Giacalone2000,Chollet2009,Chollet2011,Guo2014}. However, the particle can still move transverse to the mean magnetic field through the field line meandering \citep{Giacalone2000,Guo2014}. This perpendicular diffusion is likely required to explain large longitudinal spread of SEPs \citep{Dresing2012,Giacalone2012,Wiedenbeck2013,Zhao2018}. While these physical processes have been known, their observational consequence has not been fully explored.

In this paper, we propose a model to understand the recent SEP events observed by PSP. The model solves the propagation of SEPs instantaneously released at compact sources in synthetic solar wind magnetic turbulence \citep{Matthaeus1990,Giacalone1999,Guo2014}. We find that when the spatial locations of source regions of different species (H, $^4$He, $^3$He) are offset by a distance comparable to the radii of the source regions, the observed energetic particle composition at different locations in interplanetary space can vary significantly from the average compositions at the combined source region. This spatial offset is more established for energetic protons versus energetic electrons, as indicated by X-ray and $\gamma$-ray observations \citep{Hurford2003,Hurford2006} and interplanetary energetic particle observation \citep{Chollet2009}, but may exist between other species as well. These compositional variations are stronger when observed closer to the Sun, making PSP and SO unique for studying them more systematically in the future. The rest of this Letter is organized as follows: Section 2 briefly introduces the numerical model and parameters we use. Section 3 discusses simulation results especially the effects of turbulent magnetic fields on the slightly offset ion sources. More discussion and further implications on future observations closer to the Sun by PSP and SO are presented in Section 4.

\section{Numerical Simulations for Charged Particle Propagation in the interplanetary turbulence\label{sec:simulation}}

We consider the propagation of SEPs (protons, $^4$He, and $^3$He) released from spatially compact and instantaneous sources in turbulent solar wind magnetic fields.
We numerically integrate gyromotions of an ensemble of energetic charged particles in an advecting, synthetic turbulent magnetic field similar to that in the solar wind. 
The method we use is similar to our earlier work \citep{Giacalone1999,Guo2014,Guo2015}. Here we lay out the salient details of the model. We consider the propagation of SEPs from source regions with small spatial offsets, as illustrated in Fig. 1 a). For simplicity we only consider one turbulence model, namely the so-called composite turbulence model \citep{Matthaeus1990}, which has been often used to study energetic particle transport \citep[e.g.,][]{Giacalone1999,Qin2002}. While some detailed differences between different turbulence models have been noted before \citep{Guo2014}, they are not particularly important for the purpose of the current study, and therefore we defer a comparison among different turbulence models to a future work. 

\subsection{Turbulent Interplanetary Magnetic Field \label{sec:setup}}

The turbulent magnetic field in our model is in a three-dimensional Cartesian geometry ($x, y, z$)
\begin{eqnarray}
    \textbf{B} &=& \textbf{B}_0 + \delta \textbf{B}  %\\
    %& = & B_0 \hat{z} + \delta B_x(x, y, z, t)\hat{x} + \delta B_y(x, y, z, t)\hat{y} + \delta B_z(x, y, z, t)\hat{z} \nonumber
\end{eqnarray}

The fluctuating component is generated by summing over a range of plane wave modes with a randomly distributed propagation direction, and with randomized polarizations and phases. This is given by 

\begin{eqnarray}
     \delta\textbf{B}&=&\sum^{N_m}_{n=1}A_n \hat{\xi}_n \exp(ik_nz_n'+i\beta_n),
\end{eqnarray}

where

\begin{eqnarray}
    \hat{\xi}_n = \cos \alpha_n \hat{x}_n' + i\sin \alpha_n \hat{y}_n',
\end{eqnarray}

and 

 \begin{equation*}
    \begin{pmatrix}
      \begin{matrix}
        x' \\
        y' \\
        z'
      \end{matrix} 
      \end{pmatrix} = 
            \begin{pmatrix}
            \begin{matrix}
        \cos \theta_n \cos \phi_n & \cos \theta_n \sin \phi_n & -\sin \theta_n \\
        - \sin \phi_n & \cos \phi_n & 0 \\
         \sin \theta_n \cos \phi_n & \sin \theta_n \sin \phi_n & \cos \theta_n
      \end{matrix}
    \end{pmatrix}  
                \begin{pmatrix}
            \begin{matrix}
        x \\
        y \\
        z
      \end{matrix}
    \end{pmatrix}.  
  \end{equation*}

\noindent where $\beta_n$ is the phase of each wave mode $n$ with wavenumber $k_n$, $A_n$ is its amplitude, $\alpha_n$ is the polarization angle, and $\phi_n$ and $\theta_n$ determine the direction of the $k$-vector. $\beta_n$, $\alpha_n$, and $\phi_n$ are random numbers between $0$ and $2\pi$.
This assumes a globally uniform background magnetic field $\textbf{B}_0$ along a certain direction and a fluctuating magnetic field component $\delta \textbf{B}$. We assume the background magnetic field is along the $z$ direction. We use the composite model, which is a magnetostatic model for the wave-vector spectrum of magnetic fluctuations based on observations of the solar wind turbulence \citep{Matthaeus1990}. The fluctuating magnetic field in the model is expressed as the sum of two parts: a slab component $\delta B^s = (B_x^s(z), B_y^s(z), 0)$ and a two-dimensional component  $\delta B^{2D} = (B_x^{2D}(x, y), B_y^{2D}(x, y), 0)$. The two-dimensional component only consists
of magnetic fluctuation with wave vectors along the transverse direction $x$ and $y$ ($\theta_n = \pi/2$ and
$\alpha_n = \pi/2$), whereas the slab component is a
one-dimensional magnetic fluctuation where all wave vectors are along the direction
of the uniform magnetic field $z$ ($\theta_n=0$). Solar wind observations have suggested that the magnetic field fluctuation has components with wave
vectors either nearly parallel or perpendicular to the magnetic field, and more wave power is
concentrated in the perpendicular directions (about $80\%$) \citep{Bieber1996}. In the model we consider, the wave propagation is not included and the time dependence is only through the solar wind advection as the magnetic field is frozen in the background plasma.
This model captures the anisotropic characteristic of the solar wind turbulence. We construct the magnetic fluctuations, using the random phase approximation and assuming a Kolmogorov-like magnetic power spectrum approximated by a large number of wave modes. The slab component $\delta \textbf{B}^s$ and two-dimensional component $\delta \textbf{B}^{2D}$ can be expressed as the real parts of the following equations: 

\begin{eqnarray}
    \delta\textbf{B}^s&=&\sum^{N_m}_{n=1}A_n[\cos\alpha_n(\cos \phi_n\hat{x}+\sin\phi_n\hat{y})\\
    & &+i\sin\alpha_n(-\sin\phi_n\hat{x}+\cos\phi_n\hat{y})]\nonumber \\
    & &\times\exp(ik_nz+i\beta_n),\nonumber
\end{eqnarray}
and, 
\begin{eqnarray}
    \delta\textbf{B}^{2D}&=&\sum^{N_m}_{n=1}A_ni(-\sin \phi_n \hat{x}+\cos\phi_n\hat{y})  \\
    & &\times\exp[ik_n(\cos\phi_nx+\sin\phi_n y)+i\beta_n],\nonumber
\end{eqnarray}

The amplitude of magnetic fluctuation at wave number $k_n$ follows a Kolomogorov-like power law: 

\begin{eqnarray}
    A_n^2=\sigma^2\frac{\Delta V_n}{1+(k_nL_c)^\gamma}\left[\sum^{N_m}_{n=1}\frac{\Delta V_n}{1+(k_nL_c)^\gamma}\right]^{-1},
\end{eqnarray}

\noindent where $\sigma^2 = \delta B^2/B_0^2$ is the magnetic wave variance and $\Delta V$ is a normalization factor $\Delta V = \Delta k_n$ for the 1D slab fluctuations and $\Delta V = 2 \pi k_n \Delta k_n$ for the 2D fluctuations. $\gamma = 5/3$ for the slab component and $8/3$ for the 2D component, respectively. A logarithmic spacing in wavenumbers $k_n$ is chosen so that $\Delta k_n/k_n$ is a constant.
In our simulations, we use parameters similar to what is observed in the solar wind. The correlation length is taken to be $L_c = 0.01$ AU. The maximum and minimum wavelengths used to generate the turbulence are $\lambda_{max} = 0.3$ AU and $\lambda_{min} = 5\times 10^{-5}$ AU, respectively. We choose $\Delta k_n/k_n=0.02$.  These include the full scale of magnetic fluctuation from the scales larger than $L_c$ down to the scales that resonantly interact with particles. The variance of turbulence is taken to be $\sigma^2  = 0.15$, and, after the above implementation we find the constructed fluctuation has a variance slightly differs from the assigned value by several percent. %$k_{res} \sim |\Omega/v \mu|$, where $\Omega$ is the gyrorperiod, $\mu$ is the pitch-angle for energetic particles, and $v$ is the energetic particle speed (see below).  
The mean magnetic field $B_0$ is taken to be $15 $ nT, consistent with observations at PSP at 0.3 AU. The solar wind speed is set to be $400$ km/s. These parameters are taken to not vary with distance from the Sun because of our assumed geometry. Since the model assumes a uniform solar wind speed in Cartesian coordinates, it does not include the effects of an expanding solar wind such as adiabatic cooling, adiabatic focusing, and drift motion in a Parker spiral magnetic field \citep[e.g.,][]{Dalla2017}. Nevertheless, the model is sufficient to study the proposed physical processes on the ion composition in SEP events. In the future, it would be interesting to consider more realistic geometry and spatial dependence of background solar-wind plasma parameters, magnetic field and their fluctuations, as they may vary considerably in latitudinal and lonitudinal directions \citep[e.g.,][]{Giacalone2000,Zhao2020}.

\subsection{Energetic Particle Motions}
In the simulations, energetic particles are injected in source regions at $z = 0$.  This is done by initializing upward propagating particles $v_z > 0$ with a randomized pitch angle and gyrophase. The released particles follow a power-law velocity distribution with $f \propto v^{-4}$ corresponding to energies from $0.5$ MeV/nucleon to $100$ MeV/nucleon. After the injection the equations of motions of particles are numerically integrated using the Burlirsh-Stoer method \citep{Press1986}. The algorithm uses an adaptive time step based on an evaluation of the local truncation error. It is very efficient when the fields are fairly smooth compared with the particle gyroradius. The energy is conserved to a high accuracy, with root of the mean square difference of the total energy ranges from 0.045\% (0.5 MeV/nucleon) to 0.01\% (100 MeV/nucleon) during the simulation with no solar wind flow (zero electric field). The charged particles are released impulsively and their trajectories are numerically integrated until they reach boundaries at $z=0.5$ AU and $z=-0.03$ AU. The spacecraft observations at $0.3$ AU are mimicked by collecting particles in windows of a dimension $L_x \times L_y = 0.01 \times 0.01$ AU$^2$ when the particles pass the windows at $z=0.3$ AU. These collection regions are uniformly distributed in the x and y directions ($-0.5$ AU < x < $0.5$ AU, $-0.5$ AU < y < $0.5$ AU). The size of the collection regions is about the same size as a correlation length $L_c$, which is smaller than the typical scale of the field line random walk when the distance is much larger than $L_c$ \citep[e.g.,][]{Giacalone1999}. We have also tested smaller collection windows and the results are consistent with those presented in this paper. We also include a case where particles are tracked until observers at 1 AU to study the radial evolution of ion composition. The collected particle energies and arrival times in each window are analyzed as
a spacecraft observation. We release different ion species (H, $^4$He, and $^3$He) from the source regions ($^4$He$-$H simulations are made separately from $^3$He$-^4$He simulations). The source regions are initially chosen to be
circles at the $z=0$ plane with a radius of $R_{0} = 0.2L_c$ and the offset distance in the x direction $D_{0} = R_0 =  0.2L_c$, is smaller than
the correlation scale. Considering that the supergranulation motion scale at the Sun is $\sim 30,000$ km and roughly corresponds to the correlation length \citep{Rincon2018}, this approximately corresponds to a source region of a diameter about $12,000$ km on the Sun. We study how the offset distance and source size influence the observations at $0.3$ AU, similar to what PSP can observe. We examine different offset distances $D = 0$, $0.5 R_0$, and $2.0 R_0$ and different source sizes $R_{source} = 0.5 R_0$, $2.0 R_0$ and $4.0 R_0$. %To quantify the effect of nonuniform injection, we also examine cases with different background source $10 \%$, $20 \%$, and $50 \%$ of the particle numbers in the compact source in a larger radius 5 times of the compact source radius.  
For each run, we release about $8$ million particles for each species to get a sufficient number of particles collected in the windows and then the ion compositions of different species are calculated by assuming the composition ratios at the combined source region (He/H)$_n = 0.036$ and ($^3$He/$^4$He)$_n$ = 0.1. To ensure good statistics in each observation sample, we only examine windows that have more than $1000$ protons. Each parameter set is simulated using $4$ different realizations by utilizing a different set of random parameters. For the run with distance to 1 AU, we increased the total particle number to $80$ million (assuming the SEP intensity decays as a function of distance).

\section{Simulation Results \label{sec:results}}

\begin{figure*}
\centering
\includegraphics[width=0.8\textwidth]{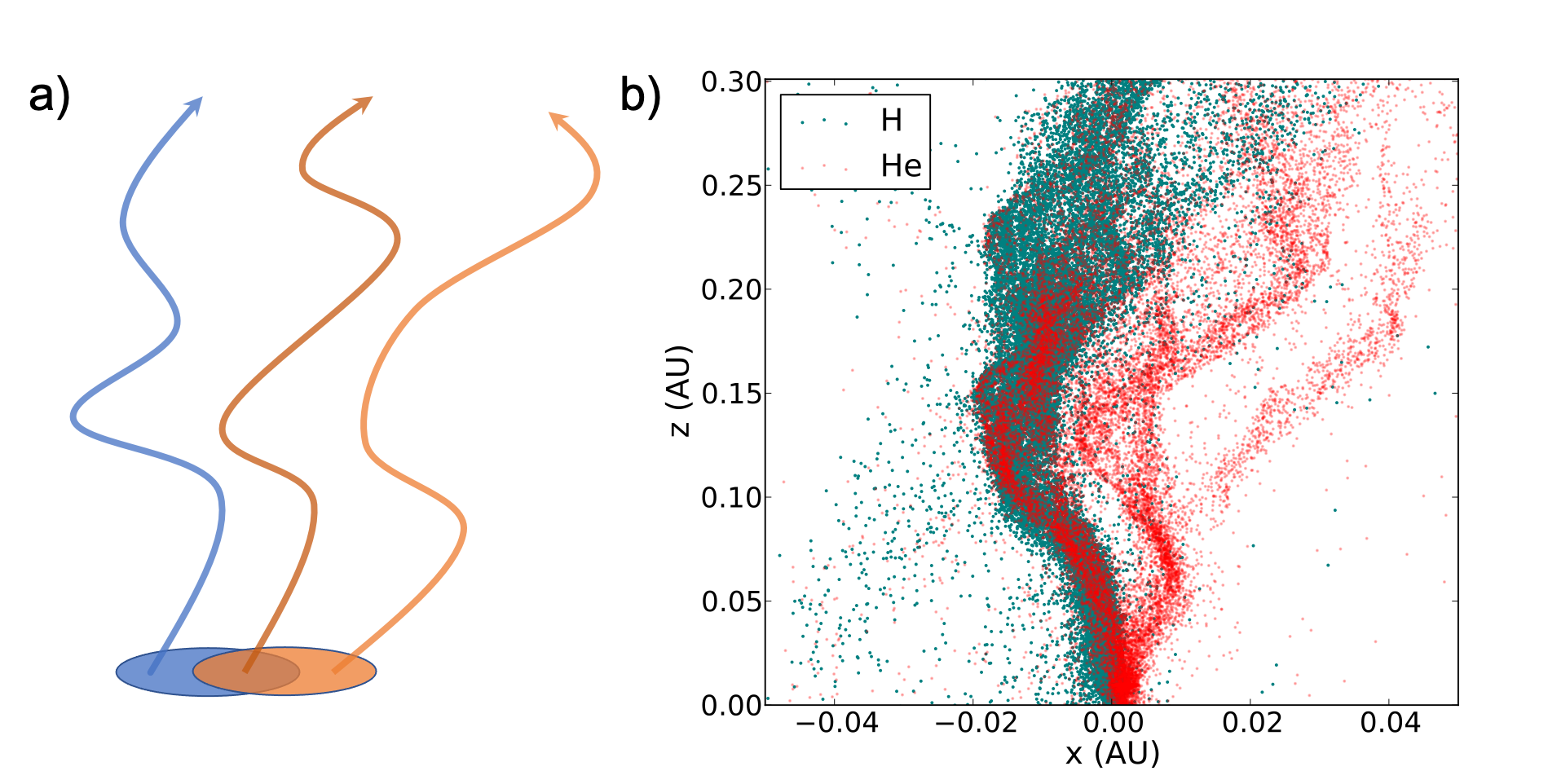}
\caption{Illustration of different field lines in a turbulent solar wind and the energetic particle distribution in the turbulent magnetic fields. a) A sketch of different field lines from two source regions with an offset. b) The distribution of energetic hydrogen (turquoise) and helium (red) after 1 hour. Observers at different locations naturally see quite different He/H ratios.}
\label{fig:distribution}
\end{figure*}

The turbulent, braiding magnetic field lines lead to complicated energetic particle distributions. 
Figure \ref{fig:distribution} a) illustrates several sample field lines from two source regions with an offset. In this case the source size is $R_{source} = R_0$ and the offset distance is $D = R_0$.    The blue source is only filled with protons and the orange source region only has helium. Particles have a tendency to move along a field line for a distance before jumping to another. Therefore the observers connected to the blue field line mostly see protons and the ones connected to the orange field line see more helium. The brown field line has a mixture of both as it connects to the overlapping region. The simulation results confirm this basic idea. 
Figure \ref{fig:distribution} b) shows the spatial distribution of protons (turquoise) and helium (red) 1 hour after the release time from the offset source. Note our magnetic field model is fully 3D, which is important to  model particle transport more realistically \citep{Jokipii1993,Jones1998}. The distributions of SEPs are highly nonuniform due to the effects of large-scale field line random walk. With this offset between the source regions, the difference between distributions of hydrogen and helium is quite significant. Spacecraft at 0.3 AU naturally see different He/H at different locations.

\begin{figure*}
\centering
\includegraphics[width=\textwidth]{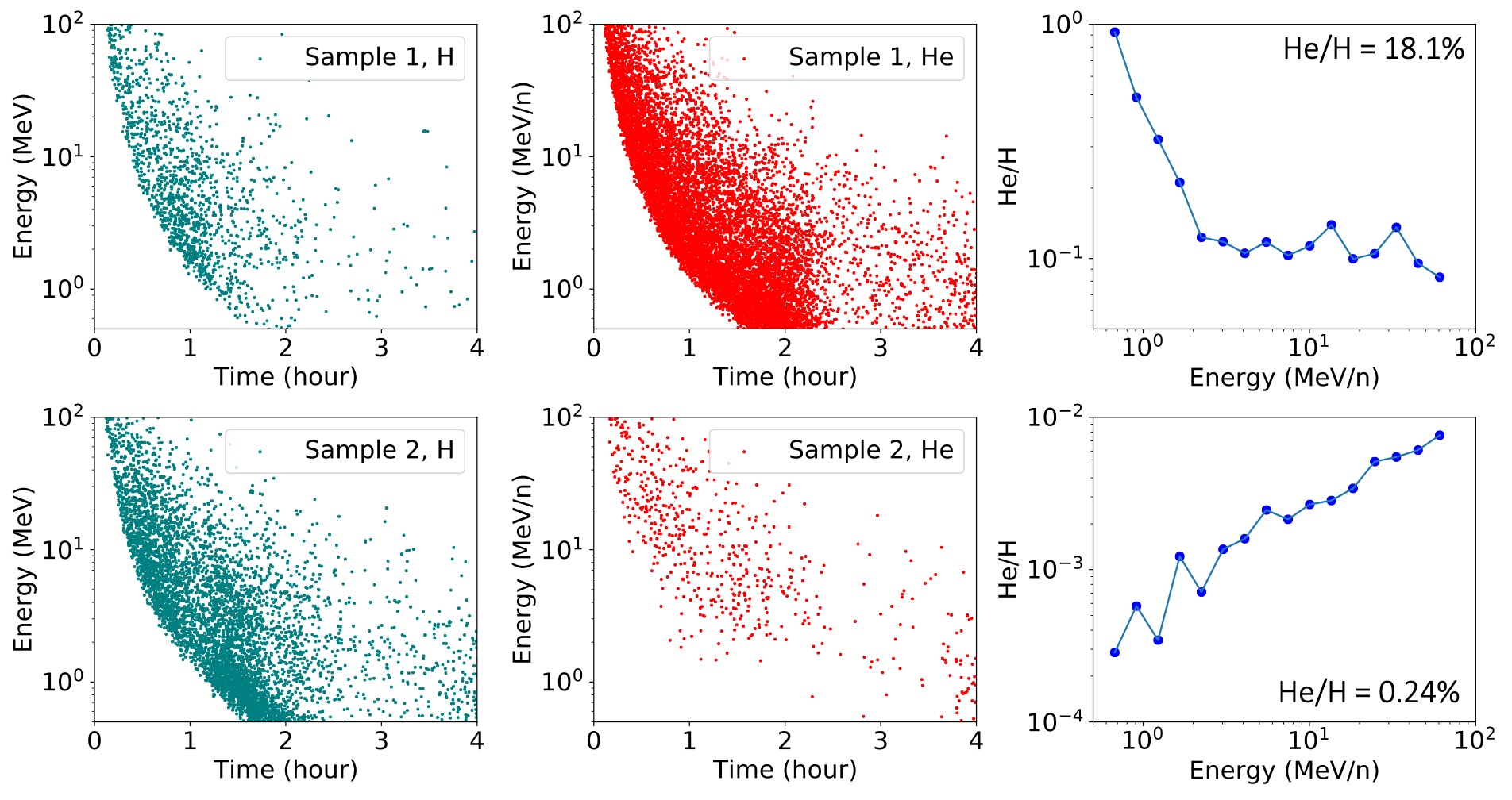}
\caption{Two sample observations taken at different locations at 0.3 AU. Upper panels show a case with low He/H = 0.181 and the lower panels show a sample event with high He/H = 0.0024. Note that the ratio has been normalized by the assumed nominal ratio (He/H)$_n$ = 0.036 (each helium count corresponds to 0.036 helium particle).}
\label{fig:xraysim}
\end{figure*}

%Due to the offset in the source regions, different observers in the inner heliosphere would see very different ion composition ratios.
Figure 2 shows two sample observations at 0.3 AU mimicking PSP observations taken at different collection locations. The upper panels show protons and helium for a sample observation that is ``helium-rich'' (He/H = 0.181), whereas the lower panels are for a helium-poor sample observation (He/H = 0.0024).  In both sample events, we observe clear velocity dispersion as faster (high energy) particles arrive at observers earlier. Sample 1 is more than 5 times of the nominal value of He/H and Sample 2 is only 1/15 of the nominal ratio used in this study. These are similar to the variation of He/H of SEPs observed by PSP \citep{Cohen2021}. We also briefly examined how the observed He/H depends on particle energy. While the energy-dependent He/H during individual observations can be quite variable, during these ``extreme'' events with very high or very low He/H ratios, they often shows clear energy dependence. At low energy, He/H shows extreme values (nearly 1.0 in Sample 1 and less than $10^{-3}$ in Sample 2), likely due to a good magnetic connection with one of the source regions and the poor connectivity with the other one. However, high energy particles can travel across field lines more easily and mix, therefore He/H in high energy has a trend to be closer to the nominal ratio (3.6 \% in this case) than that of lower energies. We also observed other sample events with different path lengths and dropouts similar to our earlier work \citep{Guo2014}, which might be seen by PSP as more SEP events are collected. Note that this variation  can exist between any two species such as $^3$He/$^4$He (see below), as long as the source regions have an offset.

\begin{figure*}
\centering
\includegraphics[width=0.9\textwidth]{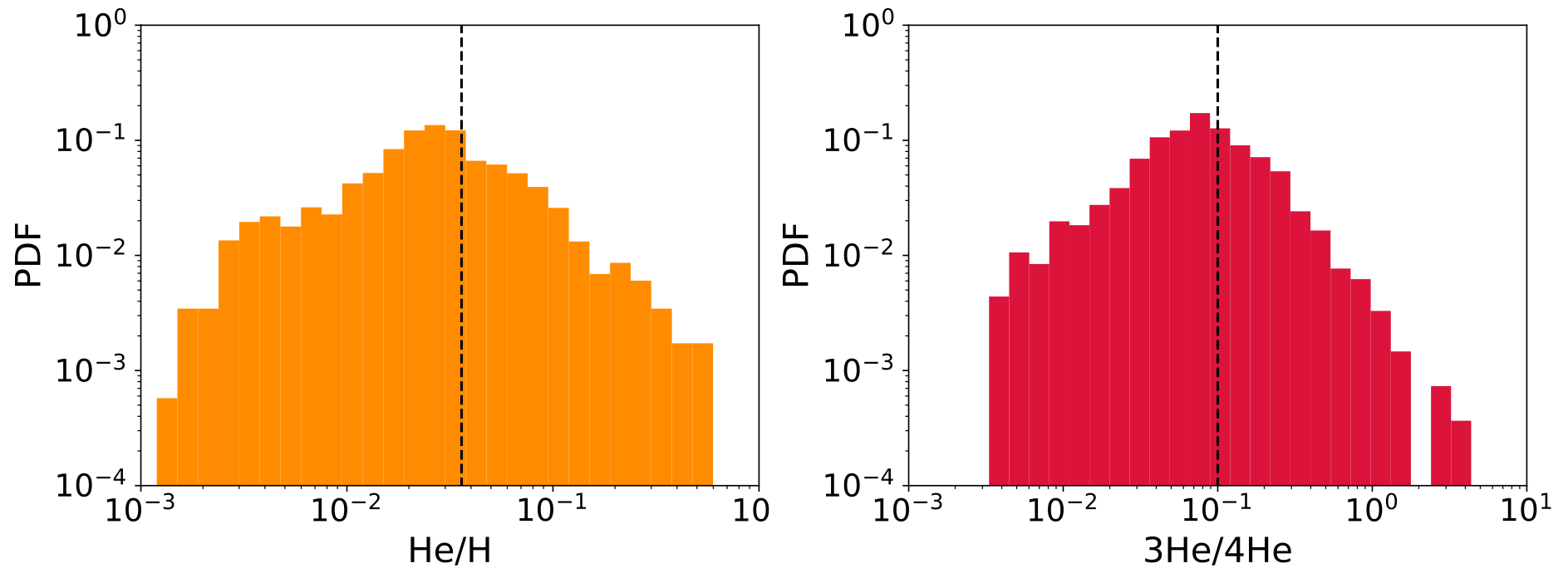}
\caption{PDFs of He/H and $^3$He/$^4$He observed at different locations at 0.3 AU when $R = R_0$ and $D=R_0$. Vertical dashed lines represent the input source ratios.}
\label{fig:histogram}
\end{figure*}

Since the observations would highly depend on how braiding magnetic field lines connect with observers, it is interesting to study the likelihood of ion composition ratios observed by spacecraft. 
We statistically study the He/H ratios and also $^3$He/$^4$He with $D = R_0$ and $R_{source} = R_0$. 
Figure 3 shows the probability distribution functions (PDFs) for He/H and $^3$He/$^4$He with more than 1000 samples collected by different observers at 0.3 AU. One can observe strong influence of interplanetary meandering magnetic fields on He/H. He/H is seen to be variable over about nearly three orders of magnitude from about $0.1\%$ to about $50\%$.  Interestingly, $^3$He/$^4$He (assuming a source ratio 0.1) shows a similar variability ranging from $0.3\%$ to about $5$. These may have strong implications on observations of ion compositions in the inner heliosphere. It is worth noting that the distribution prefers helium-poor events (He/H < (He/H)$_n$) over helium-rich events (He/H > (He/H)$_n$), which seems to be a robust feature in the simulations we have done.  The same situation applies to the $^3$He/$^4$He ratio as well. %simply because the total number of helium particles is finite and it is more difficult to produce helium-rich events in randomized samples. 

\begin{figure*}
\centering
\includegraphics[width=0.5\textwidth]{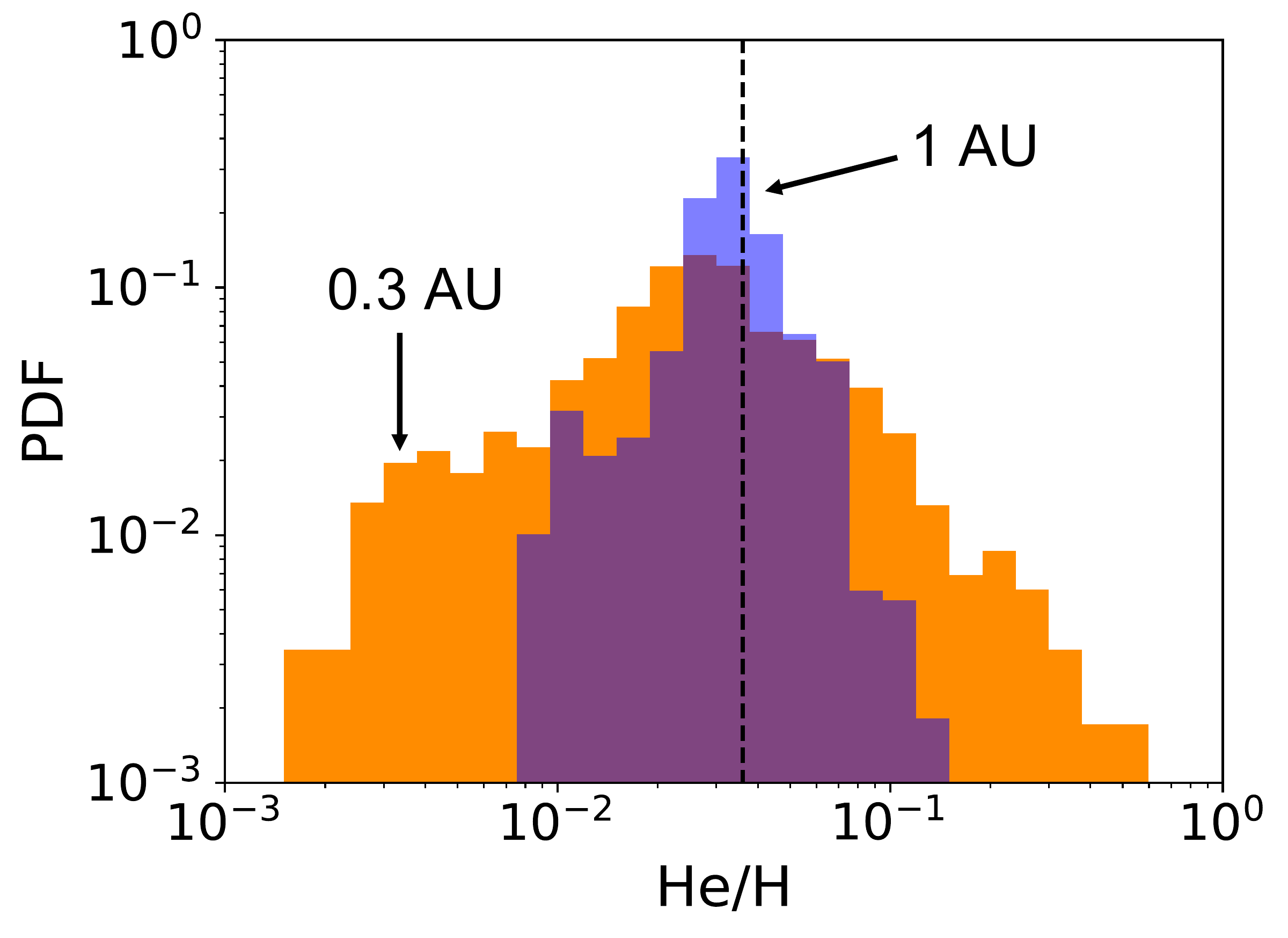}
\caption{Comparison between PDFs of He/H measured at 0.3 AU (orange) and 1 AU (blue) with a vertical line showing the source region ratio.}
\label{fig:distance}
\end{figure*}

It is instructive to explore how the distance between observers and the source regions influence the composition ratio. Figure \ref{fig:distance} shows the PDFs of He/H ratio seen at 0.3 AU and 1 AU. Samples at 1 AU still show a broad distribution for over more than one order of magnitude, but the distribution is significantly narrower than that measured at 0.3 AU. As energetic particles propagate, they are more and more mixed together even if their composition is very nonuniform in the beginning. This suggests that the variation of the composition ratio decays with distance.  The PDF at 1 AU is also qualitatively similar to the variable He/C distribution obtained by \citet{Reames2019}.

\begin{figure*}
\centering
\includegraphics[width=\textwidth]{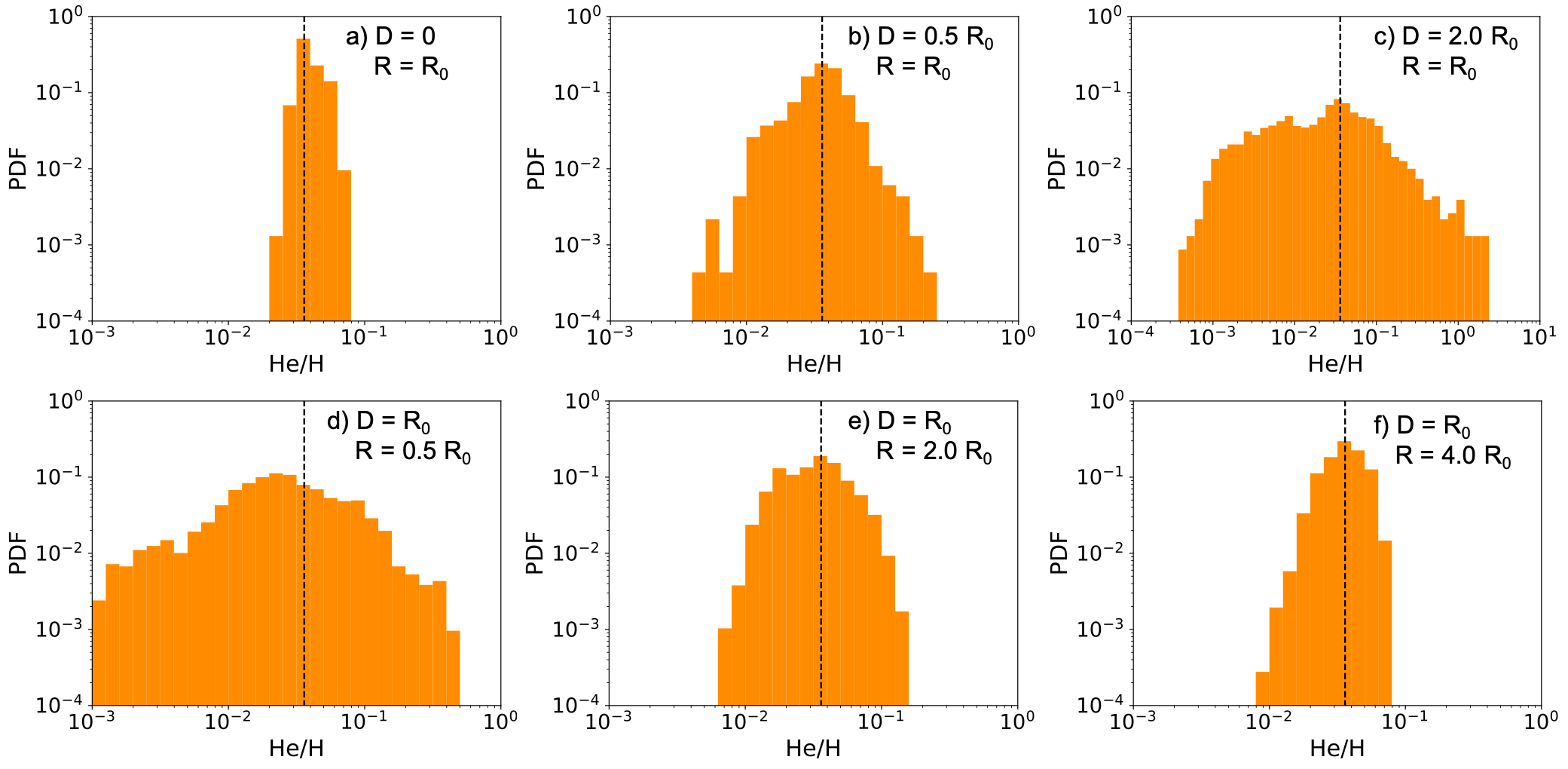}
\caption{PDFs of He/H with different displacement and source radius. Vertical lines show the assumed source region ratio $3.6 \%$. a), b), and c) show distribution of He/H for source radius $R=R_0$ and offset distance $D = 0$, $0.5 R_0$, and $2.0 R_0$, respectively. d), e), and f) show PDFs of He/H for offset distance $D = R_0$, and source radius $R= 0.5 R_0$, $2.0 R_0$, and $4.0R_0$, respectively.}
\label{fig:hist}
\end{figure*}

Next we examine how the distribution of He/H changes with offset distances (D = 0, 0.5$R_0$, 2.0$R_0$) with fixed source radii $R = R_0$, as well as different source radius (R = 0.5$R_0$, 2.0$R_0$, 4.0$R_0$) with a fixed offset distance $D = R_0$. Since PSP observations have provided quantitative measurements for He/H, and $^3$He/$^4$He is likely more variable in the combined source region \citep{Petrosian2004}, we focus on He/H, but note that $^3$He/$^4$He would have the same level of variation even for a fixed composition ratio in the combined source regions.
Figure \ref{fig:hist} shows the PDF for different cases. In the case with no displacement, the He/H is less than a factor of $2-3$ above or below the source ratio, which may due to, for example, different particle drift motions in the turbulent magnetic field. This variation becomes higher when the displacement is larger. Over all, the variation of composition ratios becomes significant when the offset distance is comparable to the radius of the source region. We do notice that the PDFs of He/H have some difference when both $R$ and $D$ change by the same ratio, indicating that the variation of isotope ratios have some weak dependence on other factors, which we plan to explore more in the future.  

%Figure 5 shows the case with background and larger source regions.

%\begin{figure*}
%\centering
%\includegraphics[width=0.3\textwidth]{HeH-back005.pdf}
%\includegraphics[width=0.3\textwidth]{HeH-back02.pdf} 
%\includegraphics[width=0.3\textwidth]{HeH-back05.pdf}
%\caption{Histograms of He/H with background.}
%\label{fig:distribution}
%\end{figure*}

\section{Discussion and Conclusions \label{sec:discussion}}

Recent PSP observations within 0.5 AU suggest substantial variations in the ion composition of SEP events \citep{Cohen2021,Leske2020,Wiedenbeck2020}. We have shown in this paper that this variation can possibly be due to the propagation of SEPs in braiding interplanetary magnetic field lines of force, if the source regions are offset. 
%Although we do not fully understand this source offset, it is likely that the offset is not going to increase for larger sources, and therefore this effect may prefer small events.
This mechanism predicts that observers at different locations will see different ion composition ratios from the same SEP event, with possible energy dependences. The variation of composition ratios decays with radial distance, which can also be tested by spacecraft observations. 
The active observations by PSP, SO and other spacecraft will likely advance our knowledge on this issue by measuring events at distances that have never been seen before. 

Earlier work has shown FIP effects may lead to a reduced He/H in the source region ratios by a factor of a few \citep{Reames2019,Laming2021}, but may have difficulties to reach extremely low He/H. Heating of helium particles can be more efficient than that of protons during magnetic reconnection in the presence of a strong guide field, potentially enhancing He/H in the energetic component \citep{Drake2014}.
Our model suggests that interplanetary turbulent magnetic fields can lead to significant variability both above and below the nominal ratios, without needing a mechanism to change the source ratios. 
  Since the simulated variation in our paper is biased toward smaller He/H than the nominal values, it may be often seen as a suppression \citep{Reames2019}. In the future, it would be important to statistically study the variable composition ratio over a large number of species in many SEP events \citep[e.g.,][]{Wiedenbeck2008,Bucik2021}.  

%Although we do not know the full details of this, it is possible that some small events have been diffuse away when particles arrive at 1 AU, making the measurements incomplete. 

%It has been known that the electron to proton ratios are highly variable. 

%We also find dropout features in our simulations (not shown) \citep{Mazur2000,Giacalone2000,Ruffolo2003,Chollet2008,Guo2014}.
PSP has reported dropout-like features where proton flux suddenly drops \citep{Leske2020} similar to that observed at 1 AU \citep{Mazur2000,Giacalone2000,Ruffolo2003,Chollet2008,Guo2014}. If only one species suddenly changes its flux, this may lead to a temporal change of He/H.  If the helium has a similar dropout phenomenon with a different timing, this can be used to test whether helium and hydrogen source regions are offset. \citet{Chollet2009} have shown some electron dropouts have different timings from the proton dropouts which may be due to the separation of the source region. In addition, previous observations have demonstrated that different observers will see different characteristics of SEP features \citep{Chollet2008}. Future PSP and SO observations may provide further insight on this.

We have shown that in our model the source region offset is essential for producing the observed variations in ion compositions by PSP. Thus, on the one hand, our results suggest that observations of composition variations observed {\it in situ} provide important information about variations in the combined source region close to the Sun. On the other hand, however, it is unclear what leads to this offset. There have been observations showing the spatial offsets between electrons (indicated by X-rays) and ions (indicated by $\gamma$-rays). 
The observed offset in several flare events is about 10,000-15,000 km \citep{Lin2003,Hurford2003,Hurford2006}, which is comparable to the offset distance needed to produce the observed composition variations reported. The mechanism for this offset is not fully understood. \citet{Emslie2004} suggest that this offset is due to the fact that different particle species will interact with MHD turbulence differently during stochastic acceleration process \citep{Petrosian2004,Fu2020}. Meanwhile, there have been observations and theoretical work showing protons and electrons can be accelerated simultaneously by the same mechanism \citep{Shih2009,Guo2012,Zhang2021}. We comment that the drift motion in the reconnection current layer may contribute to this. As an order of magnitude estimate, we note that drift motions in the current sheet scales with $v_d \sim \varepsilon/(qBR_c)$ \citep{Guo2014b,Li2017,Li2019,Arnold2021}. Assuming a magnetic field $B \sim 30$ G, curvature radius $R_c \sim 1000$ km, and particle energy $\varepsilon \sim 30 $ MeV, this can lead to a drift speed of $\sim 10$ km/s. The time scale for this drift motion may be limited by the scale of the reconnection layer and the outflow speed $\tau \sim L/V_{out}$. We assume a current sheet of $L \sim 10^5$ km and an outflow speed of $V_{out} \sim 100$ km/s \citep[e.g.,][]{Tian2014}. The resulting drift distance can be $L_d \sim 10,000$ km, similar to what was observed by RHESSI \citep{Hurford2006}, although more refined calculations are definitely needed. Finally, we note that different time-dependent injections of different species \citep{Wang2016} would also be interesting to explore in the future. 

In summary, we have presented numerical simulations for interpreting recent PSP observations on the highly variable He/H and possibly other ion composition ratios. We showed that when the source regions of different species are offset by a distance comparable to the size of the source region, the observed energetic particle composition ratios can be strongly variable, over more than two orders of magnitude. The composition variation decays with distance, making PSP and SO unique for further understanding. The observed $^3$He/$^4$He can also be quite variable if such offset exists between source regions of different species, which is consistent with statistical observations \citep{Ho2005}, although it is unknown if this variation is due to the transport effect, or entirely due to the acceleration process.

\acknowledgments{We acknowledge the constructive comments from the reviewer of this paper, which significantly helped improving the clarity of the paper. 
We thank members of SolFER collaboration for helpful feedback. SolFER is a multi-institutional collaboration for understanding the release of magnetic field energy and associated particle acceleration in flares in the solar corona. F. G., C. M. S. C., M. W., and Q. Z. acknowledge SolFER DRIVE center program with grant number 80NSSC20K0627 and 80HQTR20T0040. F. G. and L. Zhao acknowledge supports from NASA PSP GI Program 80HQTR21T0117. F. G. and J. G. acknowledge support in part from NASA LWS Program grant 80HQTR21T0005. L. Z is supported in part by NASA SOLSTICE DRIVE Science Center grant 80NSSC20K0600 and NASA LWS grant 80NSSC21K0417. The work by X. L. and F. G. is funded in part by NASA Grant 80NSSC21K1313 and 80HQTR21T0087. The work of J. G. was supported in part by NASA's Parker Solar Probe Mission under contract NNN06AA01C, NASA under grant 80NSSC20K1283, and NSF/AGS under grant 1931252. Q. Z. also acknowledges support from NASA Grant 80HQTR21T0103. F. G. is also supported by DOE through the LDRD program at LANL and office of Fusion Energy Sciences.}

%\bibliography{PIC+PolX-ray}
%\bibliographystyle{aasjournal}



\end{document}